\documentstyle[12pt]{article}
\newlength{\extraspace}
\newlength{\extraspaces}
\catcode`\@=11
\def\numberbysection{\@addtoreset{equation}{section}
\def\theequation{\arabic{section}.\arabic{equation}}}
\newcommand{\be}{\begin{equation}
\addtolength{\abovedisplayskip}{\extraspaces}
\addtolength{\belowdisplayskip}{\extraspaces}
\addtolength{\abovedisplayshortskip}{\extraspace}
\addtolength{\belowdisplayshortskip}{\extraspace}}
\newcommand{\ee}{\end{equation}}
\newcommand{\ba}{\begin{eqnarray}
\addtolength{\abovedisplayskip}{\extraspaces}
\addtolength{\belowdisplayskip}{\extraspaces}
\addtolength{\abovedisplayshortskip}{\extraspace}
\addtolength{\belowdisplayshortskip}{\extraspace}}
\newcommand{\ea}{\end{eqnarray}}
\newcommand{\newsection}[1]{
\vspace{7mm}
\pagebreak[3]
\addtocounter{section}{1}
\setcounter{subsection}{0}
\setcounter{footnote}{0}
\begin{center}
{\large {\bf \thesection. #1}}
\end{center}
\nopagebreak
\medskip
\nopagebreak
\hspace{3mm}}
\newcommand{\nonu}{\nonumber \\[.5mm]}
\newcommand{\A}{&\!\!\!}

\newcommand{\tr}{\, {\rm tr}}

\newcommand{\VEV}[1]{\left\langle {#1} \right\rangle}

\newcommand{\slA}{A \!\!\! \raisebox{0.2ex}{/}}

\begin{document}
\addtolength{\baselineskip}{.7mm}
\thispagestyle{empty}
\begin{flushright}
OCHA--PP--131 \\
STUPP--99--156 \\
{\tt hep-th/9904010} \\ 
April, 1999
\end{flushright}
\vspace{7mm}
\begin{center}
{\Large{\bf Super Weyl Anomalies in the AdS/CFT \\[2mm]
Correspondence 
}} \\[20mm] 
{\sc Madoka Nishimura}
\\[3mm]
{\it Department of Physics, Ochanomizu University \\
2-1-1, Otsuka, Bunkyo-ku, Tokyo 112-0012, Japan} \\[3mm]
and \\[3mm]
{\sc Yoshiaki Tanii}\footnote{
\tt e-mail: tanii@post.saitama-u.ac.jp} \\[3mm]
{\it Physics Department, Faculty of Science \\
Saitama University, Urawa, Saitama 338-8570, Japan} \\[20mm]
{\bf Abstract}\\[7mm]
{\parbox{13cm}{\hspace{5mm}
Anomalies of $N=(4,4)$ superconformal field theories coupled to 
a conformal supergravity background in two dimensions are computed 
by using the AdS/CFT correspondence. We find that Weyl, axial gauge 
and super Weyl transformations are anomalous, while general 
coordinate, local Lorentz, vector gauge and local super 
transformations are not. The coefficients of the anomalies show that 
the superconformal field theories have the central charge expected 
in the AdS/CFT correspondence. 
}}
\end{center}
\vfill
\newpage
\setcounter{section}{0}
\setcounter{equation}{0}
\numberbysection
%
%
\newsection{Introduction}
According to the AdS/CFT correspondence \cite{MAL,GKP,WITTEN} 
the string/M-theory in ($d+1$)-dimensional anti de Sitter (AdS) 
space times a compact space is dual to a $d$-dimensional 
conformal field theory (CFT) defined on the boundary of the AdS 
space. There is a one-to-one correspondence between boundary 
values $\phi_{0I}(x)$ of bulk fields $\phi_I(x)$ and operators 
${\cal O}_I(x)$ in the boundary CFT. The generating functional of 
correlation functions of operators ${\cal O}_I$ in the CFT is 
given by the partition function of the string/M-theory. 
In a certain limit the string/M-theory is approximated by a low 
energy effective supergravity and the partition function is given 
by a stationary point of the supergravity action $S[\phi]$ 
\be
\VEV{\exp \left(i \int d^d x \, \sum_I \phi_{0I}(x) 
{\cal O}_I(x) \right)}_{\rm CFT} 
= \exp(iS[\phi]). 
\label{adscft}
\ee
Here, $\phi_{0I}$ on the left hand side are arbitrary functions 
defined on the $d$-dimensional boundary while $\phi_I$ on the right 
hand side are the solutions of supergravity field equations 
in the bulk satisfying boundary conditions $\phi_I = \phi_{0I}$. 
\par
The boundary fields $\phi_{0I}$ are expected to form 
supermultiplets of a $d$-dimensional conformal supergravity 
\cite{FFZ,LT} since they couple to CFT operators. 
In ref.\ \cite{NT} we studied this relation between AdS 
supergravities in $d+1$ dimensions and conformal supergravities 
in $d$ dimensions in the case of three-dimensional ($p,q$) 
AdS supergravities based on the superalgebra OSp($p|$2) $\times$ 
OSp($q|$2) \cite{AT}. 
We explicitly showed that the local symmetry transformations of 
the bulk AdS supergravities induce the local transformations of 
the boundary fields $\phi_{0I}$, which coincide with those of 
two-dimensional conformal supergravities for $p,q \leq 2$. 
In particular, Weyl and super Weyl transformations on the boundary 
are generated from general coordinate and super transformations 
in the bulk. 
\par
The purpose of this paper is to compute anomalies of the 
local transformations of conformal supergravities on the boundary 
by using the AdS/CFT correspondence. By eq.\ (\ref{adscft}) the 
effective action for CFTs in the background conformal supergravity 
fields $\phi_{0I}$ is given by the classical supergravity action 
evaluated at solutions of field equations with boundary conditions 
$\phi_I = \phi_{0I}$. Anomalies are obtained by computing variations 
of the supergravity action for the local symmetry transformations 
\be
A = \delta S[\phi]. 
\label{anomaly}
\ee
Although anomalies arise as a quantum effect in the CFT side, they 
can be obtained by classical calculations in the supergravity side. 
Weyl anomaly in a purely gravitational background was computed by 
using the AdS/CFT correspondence in ref.\ \cite{HS2}. Our work is 
a generalization of this result to the case of a supergravity 
background. Anomalies of bosonic symmetries in the AdS/CFT 
correspondence were also discussed in 
refs.\ \cite{HKL,NO,GW,APTY,BK}. 
\par
We use the three-dimensional $N=(4,4)$ AdS supergravity based on the 
superalgebra SU(1,1$|$2) $\times$ SU(1,1$|$2) to compute anomalies. 
As in ref.\ \cite{NT} the local symmetry transformations in this 
bulk theory will be shown to generate those of the two-dimensional 
$N=(4,4)$ conformal supergravity \cite{PvN} on the boundary. 
The $N=(4,4)$ AdS supergravity arises from the type IIB 
supergravity on ${\rm AdS}_3 \times {\rm S}^3 \times {\rm M}_4$, 
where ${\rm M}_4$ is K3 or ${\rm T}^4$. This spacetime corresponds 
to a system of parallel $Q_1$ D1-branes and $Q_5$ D5-branes in the 
low energy limit \cite{SV,CM}. The $Q_1$ D1-strings are stretching 
in one dimension on the $Q_5$ D5-branes and the remaining 
four-dimensional directions of the D5-branes are wrapping on 
${\rm M}_4$. In the infrared limit the gauge theory on the branes 
is described by a $N=(4,4)$ super CFT with the central charge 
$c = 6 Q_1 Q_5 = {3l \over 2G}$ \cite{SV,CM}, where $l$ is the 
radius of the AdS space and $G$ is the three-dimensional 
gravitational constant. The $N=(4,4)$ AdS supergravity is expected 
to be dual to this CFT \cite{MAL,MS,BOER}. 
\par
We find that Weyl, axial gauge and super Weyl transformations are 
anomalous, while general coordinate, local Lorentz, vector gauge 
and local super transformations are not. The coefficients of the 
anomalies show that the two-dimensional CFT has the $N=(4,4)$ super 
Virasoro algebra with the central charge $c = {3l \over 2G}$ as 
expected in ref.\ \cite{MAL,MS,BOER}. This also agrees with the 
value of the central charge obtained in ref.\ \cite{BH} by 
computing the asymptotic symmetry algebra of three-dimensional AdS. 
Since our calculations of anomalies do not much depend on special 
properties in three dimensions, generalization to higher dimensions 
will be straightforward. 
\par
In the next section we give the action and the local symmetry 
transformation laws of the $N=(4,4)$ AdS supergravity in three 
dimensions. The relation to the $N=(4,4)$ conformal supergravity 
on the two-dimensional boundary is discussed in sect.\ 3. 
Boundary terms and counterterms to be added to the bulk action are 
obtained in sect.\ 4 and anomalies are computed in the final section. 
\par
\newpage
%
\newsection{Three-dimensional AdS supergravity}
We consider a three-dimensional AdS supergravity which has 
16 supersymmetries. It is based on the superalgebra 
SU(1,1$|$2) $\times$ SU(1,1$|$2), which contains 
SO(2,2) $\times$ SU(2) $\times$ SU(2) as a bosonic subalgebra. 
We call this theory as $N=(4,4)$ AdS supergravity. 
Note that this theory is different from the $N=(4,4)$ theory in 
ref.\ \cite{AT} based on the superalgebra 
OSp(4$|$2) $\times$ OSp(4$|$2), which contains 
SO(2,2) $\times$ SO(4) $\times$ SO(4) as a bosonic subalgebra. 
\par
The field content of this theory is a dreibein $e_M{}^A$, 
Rarita-Schwinger fields $\psi_M^{\alpha\dot\alpha}$, 
$\psi_M^{\alpha'\dot\alpha'}$ and SU(2) $\times$ SU(2) 
Chern-Simons gauge fields $A_M^\alpha{}_\beta 
= {1 \over 2} i (\sigma^i)^\alpha{}_\beta A_M^i$, 
$A_M^{\alpha'}{}_{\beta'} = {1 \over 2} i 
(\sigma^{i'})^{\alpha'}{}_{\beta'} A_M^{i'}$, 
where $\sigma^i$, $\sigma^{i'}$ are $2 \times 2$ Pauli matrices. 
The undotted indices $\alpha, \beta, \cdots = 1, 2$, 
$\alpha', \beta', \cdots = 1, 2$ and $i, j, \cdots = 1, 2, 3$, 
$i', j', \cdots = 1, 2, 3$ are those of the local 
SU(2) $\times$ SU(2) symmetry. 
The dotted indices $\dot\alpha, \dot\beta, \cdots = 1, 2$ and 
$\dot\alpha', \dot\beta', \cdots  1, 2$ are those of the rigid 
SU(2) $\times$ SU(2) symmetry, which is not contained in 
SU(1,1$|$2) $\times$ SU(1,1$|$2). 
We denote three-dimensional world indices as 
$M, N, \cdots = 0, 1, 2$ and local Lorentz indices as 
$A, B, \cdots = 0, 1, 2$. The Rarita-Schwinger fields satisfy 
doubly-symplectic Majorana conditions 
\be
(\psi^{\alpha\dot\alpha})^c 
= \epsilon_{\alpha\beta} \epsilon_{\dot\alpha\dot\beta} 
\psi^{\beta\dot\beta}, \qquad
(\psi^{\alpha'\dot\alpha'})^c 
= \epsilon_{\alpha'\beta'} \epsilon_{\dot\alpha'\dot\beta'} 
\psi^{\beta'\dot\beta'}, 
\label{dsm}
\ee
where the superscript $c$ denotes the charge conjugation and 
$\epsilon_{12} = - \epsilon_{21} = -1$. 
We use the following conventions. 
The flat metric is $\eta_{AB} = {\rm diag}(-1, +1, +1)$ and 
the totally antisymmetric tensor $\epsilon^{ABC}$ is chosen as 
$\epsilon^{012} = +1$. $2 \times 2$ gamma matrices $\gamma_A$ 
satisfy $\{\gamma_A, \gamma_B \} = 2 \eta_{AB}$. 
$\gamma$'s with multiple indices are antisymmetrized products 
of gamma matrices with unit strength. In particular, we have 
$\gamma^{ABC} = - \epsilon^{ABC}$ in three dimensions. 
The Dirac conjugate of a spinor $\psi$ 
is defined as $\bar\psi = \psi^\dagger i \gamma^0$. 
\par
The action consists of three parts: 
\be
S = S_{\rm bulk} + S_{\rm boundary} + \Delta S, 
\label{totalaction}
\ee
where $S_{\rm bulk}$ is the bulk action while 
$S_{\rm boundary}$ and $\Delta S$ represent boundary terms 
and counterterms respectively, which will be discussed 
in sect.\ 4. The bulk action is given by 
\ba
S_{\rm bulk} 
\A = \A {1 \over 8\pi G} \int d^3 x \biggl[ \, 
{1 \over 2} e R + 4 m^2 e \nonu
\A \A + {1 \over 2} i \epsilon^{MNP} \bar\psi_{M\alpha\dot\alpha} 
{\cal D}_N \psi_P^{\alpha\dot\alpha} 
+ {1 \over 2} i m e \bar\psi_{M\alpha\dot\alpha} \gamma^{MN} 
\psi_N^{\alpha\dot\alpha} \nonu
\A \A + {1 \over 2} i \epsilon^{MNP} \bar\psi_{M\alpha'\dot\alpha'} 
{\cal D}_N \psi_P^{\alpha'\dot\alpha'} 
- {1 \over 2} i m e \bar\psi_{M\alpha'\dot\alpha'} 
\gamma^{MN} \psi_N^{\alpha'\dot\alpha'} \nonu
\A \A - {1 \over 4m} \epsilon^{MNP} \tr \left( 
A_M \partial_N A_P + {2 \over 3} A_M A_N A_P \right) \nonu
\A \A + {1 \over 4m} \epsilon^{MNP} \tr \left( 
A'_M \partial_N A'_P + {2 \over 3} A'_M A'_N A'_P \right) 
\, \biggr], 
\label{bulkaction}
\ea
where $m$ is a positive constant and is related to the radius of 
the AdS space $l$ as $l = {1 \over 2m}$. In the following the 
gravitational constant will be chosen as $8\pi G = 1$. 
Our conventions for the curvature tensors and the covariant 
derivatives are 
\ba
R \A = \A e_A{}^M e_B{}^N R_{MN}{}^{AB}, \nonu
R_{MN}{}^{AB} \A = \A \partial_M \omega_N{}^{AB} 
+ \omega_M{}^A{}_C \omega_N{}^{CB} - (M \leftrightarrow N), \nonu
{\cal D}_M \psi_N^{\alpha\dot\alpha} 
\A = \A \left( \partial_M + {1 \over 4} \omega_M{}^{AB} 
\gamma_{AB} \right) \psi_N^{\alpha\dot\alpha} 
+ A_M^\alpha{}_\beta \psi_N^{\beta\dot\alpha}, \nonu
{\cal D}_M \psi_N^{\alpha'\dot\alpha'} 
\A = \A \left( \partial_M + {1 \over 4} \omega_M{}^{AB} 
\gamma_{AB} \right) \psi_N^{\alpha'\dot\alpha'} 
+ A_M^{\alpha'}{}_{\beta'} \psi_N^{\beta'\dot\alpha'}. 
\ea
The covariant derivatives without SU(2) $\times$ SU(2) 
connection terms are denoted as $D_M$. 
The spin connection $\omega_M{}^{AB}$ is determined 
by the torsion condition 
\be
D_M e_N{}^A - D_N e_M{}^A 
= {1 \over 2} i ( 
\bar\psi_{M\alpha\dot\alpha} \gamma^A \psi_N^{\alpha\dot\alpha} 
+ \bar\psi_{M\alpha'\dot\alpha'} \gamma^A 
\psi_N^{\alpha'\dot\alpha'} ). 
\ee
\par
The bulk action (\ref{bulkaction}) is invariant under the 
following local transformations up to total derivative terms: 
\ba
\delta e_M{}^A 
\A = \A \xi^N \partial_N e_M{}^A + \partial_M \xi^N e_N{}^A 
- \lambda^A{}_B e_M{}^B 
+ {1 \over 2} i \left( \bar\epsilon_{\alpha\dot\alpha} 
\gamma^A \psi_M^{\alpha\dot\alpha} 
+ \bar\epsilon_{\alpha'\dot\alpha'} \gamma^A 
\psi_M^{\alpha'\dot\alpha'} \right), \nonu
\delta \psi_M^{\alpha\dot\alpha} 
\A = \A \xi^N \partial_N \psi_M^{\alpha\dot\alpha} 
+ \partial_M \xi^N \psi_N^{\alpha\dot\alpha} 
- {1 \over 4} \lambda^{AB} \gamma_{AB} \psi_M^{\alpha\dot\alpha} 
- \theta^\alpha{}_\beta \psi_M^{\beta\dot\alpha} 
+ {\cal D}_M \epsilon^{\alpha\dot\alpha} 
+ m \gamma_M \epsilon^{\alpha\dot\alpha}, \nonu
\delta \psi_M^{\alpha'\dot\alpha'} 
\A = \A \xi^N \partial_N \psi_M^{\alpha'\dot\alpha'} 
+ \partial_M \xi^N \psi_N^{\alpha'\dot\alpha'} 
- {1 \over 4} \lambda^{AB} \gamma_{AB} \psi_M^{\alpha'\dot\alpha'} 
- \theta^{\alpha'}{}_{\beta'} \psi_M^{\beta'\dot\alpha'} \nonu
\A \A + {\cal D}_M \epsilon^{\alpha'\dot\alpha'} 
- m \gamma_M \epsilon^{\alpha'\dot\alpha'}, \nonu
\delta A_M^i 
\A = \A \xi^N \partial_N A_M^i 
+ \partial_M \xi^N A_N^i 
+ {\cal D}_M \theta^i 
- 2 m \bar\epsilon_{\alpha\dot\alpha} \psi_M^{\beta\dot\alpha} 
(\sigma^i)^\alpha{}_\beta, \nonu
\delta A_M^{i'} 
\A = \A \xi^N \partial_N A_M^{i'} 
+ \partial_M \xi^N A_N^{i'} 
+ {\cal D}_M \theta^{i'} 
+ 2 m \bar\epsilon_{\alpha'\dot\alpha'} \psi_M^{\beta'\dot\alpha'} 
(\sigma^{i'})^{\alpha'}{}_{\beta'}. 
\label{localsym}
\ea
The transformation parameters $\xi^M(x)$, $\lambda^{AB}(x)$, 
$\theta^\alpha{}_\beta(x) 
= {1 \over 2} i (\sigma^i)^\alpha{}_\beta \theta^i(x)$, 
$\theta^{\alpha'}{}_{\beta'}(x) 
= {1 \over 2} i (\sigma^{i'})^{\alpha'}{}_{\beta'} \theta^{i'}(x)$, 
and $\epsilon^{\alpha\dot\alpha}(x)$, 
$\epsilon^{\alpha'\dot\alpha'}(x)$ represent general coordinate 
($\delta_G$), local Lorentz ($\delta_L$), SU(2) $\times$ SU(2) 
gauge ($\delta_g$) and local super ($\delta_Q$) transformations 
respectively. The parameters $\epsilon^{\alpha\dot\alpha}$, 
$\epsilon^{\alpha'\dot\alpha'}$ satisfy the doubly-symplectic 
Majorana conditions (\ref{dsm}). 
The commutator algebra of these transformations closes 
modulo the field equations. In particular, the commutator of 
two local supertransformations with parameters $\epsilon_1$ 
and $\epsilon_2$ is 
\be
[ \delta_Q(\epsilon_1), \delta_Q(\epsilon_2) ] 
= \delta_G(\xi) + \delta_L(\lambda) 
+ \delta_g(\theta) + \delta_Q(\epsilon), 
\ee
where 
\ba
\xi^M \A = \A {1 \over 2} i \left( 
\bar\epsilon_{2\alpha\dot\alpha} \gamma^M 
\epsilon_1^{\alpha\dot\alpha} 
+ \bar\epsilon_{2\alpha'\dot\alpha'} \gamma^M 
\epsilon_1^{\alpha'\dot\alpha'} \right), \nonu
\lambda^{AB} \A = \A - \xi^M \omega_M{}^{AB} 
- i m \left( 
\bar\epsilon_{2\alpha\dot\alpha} \gamma^{AB} 
\epsilon_1^{\alpha\dot\alpha} 
- \bar\epsilon_{2\alpha'\dot\alpha'} \gamma^{AB} 
\epsilon_1^{\alpha'\dot\alpha'} \right), \nonu
\theta^i \A = \A - \xi^M A_M^i 
- 2 m \bar\epsilon_{2\alpha\dot\gamma} 
\epsilon_1^{\beta\dot\gamma} (\sigma^i)^\alpha{}_\beta, \nonu
\theta^{i'} \A = \A - \xi^M A_M^{i'} 
+ 2 m \bar\epsilon_{2\alpha'\dot\gamma'} 
\epsilon_1^{\beta'\dot\gamma'} 
(\sigma^{i'})^{\alpha'}{}_{\beta'}, \nonu
\epsilon^{\alpha\dot\alpha} 
\A = \A - \xi^M \psi_M^{\alpha\dot\alpha}, \qquad
\epsilon^{\alpha'\dot\alpha'} 
= - \xi^M \psi_M^{\alpha'\dot\alpha'}. 
\ea
The action (\ref{bulkaction}) is also invariant under the 
rigid SU(2) $\times$ SU(2) transformations, which act on 
the dotted indices of $\psi_M^{\alpha\dot\alpha}$, 
$\psi_M^{\alpha'\dot\alpha'}$. 
\par
Three-dimensional AdS supergravities with $N=(p,q)$ 
supersymmetries for $p, q = 1, 2, 4$ can be obtained from the 
above $N=(4,4)$ theory by consistent truncations. 
For instance, the $N=(2,2)$ theory in ref.\ \cite{AT} can be 
obtained by a truncation 
\be
\psi^{1\dot2} = \psi^{2\dot1} = \psi^{1'\dot2'} 
= \psi^{2'\dot1'} = 0, \quad
A_M^1 = A_M^2 = A_M^{1'} = A_M^{2'} = 0. 
\ee
Redefining the remaining fields as 
\be
\psi_M^1 \equiv {1 \over \sqrt{2}} 
( \psi_M^{1\dot1} + \psi_M^{2\dot2} ), \quad
\psi_M^2 \equiv {1 \over \sqrt{2}} i 
( \psi_M^{1\dot1} - \psi_M^{2\dot2} ), \quad
A_M^{12} = - A_M^{21} \equiv {1 \over 2} A_M^3 
\ee
and similarly for $\psi_M^{1'}$, $\psi_M^{2'}$, $A_M^{1'2'}$, 
we obtain the action and the local transformations of the 
$N=(2,2)$ theory \cite{AT}. $A_M^{12}$, $A_M^{1'2'}$ are the 
SO(2) $\times$ SO(2) gauge fields and $\psi_M^i$, $\psi_M^{i'}$ 
($i, i' = 1, 2$) are Majorana spinors. 
The $N=(1,1)$ theory \cite{AT} is obtained by further truncation 
\be
\psi_M^2 = \psi_M^{2'} = 0, \qquad
A_M^{12} = A_M^{1'2'} = 0. 
\ee
The ($p,q$) theories for other values of $p, q$ can be also 
obtained by appropriate truncations. 
\par
%
%
\newsection{Local symmetries on the boundary}
As in ref.\ \cite{NT} we partially fix the gauge for the local 
symmetries (\ref{localsym}) and obtain how the residual local 
symmetries act on the fields on the boundary. 
We represent the three-dimensional AdS space as a region 
$x^2 > 0$ in ${\bf R}^3$ with coordinates $(x^0, x^1, x^2)$. 
The boundary of the AdS space corresponds to a plane $x^2 = 0$ 
and a point $x^2 = \infty$. Our gauge fixing conditions are 
\ba
\A\A e_{M=2}{}^{A=2} = {1 \over 2mx^2}, \qquad 
e_{M=2}{}^a = 0, \qquad
e_\mu{}^{A=2} = 0, \nonu
\A\A \psi_2^{\alpha\dot\alpha} = 0, \qquad 
\psi_2^{\alpha'\dot\alpha'} = 0, \qquad
A_2^\alpha{}_\beta = 0, \qquad
A_2^{\alpha'}{}_{\beta'} = 0, 
\label{gaugefix}
\ea
where $\mu, \nu, \cdots = 0,1$ and $a, b, \cdots = 0, 1$ are 
two-dimensional world indices and local Lorentz indices 
respectively. The metric in this gauge has a form 
\be
dx^M dx^N g_{MN} = {1 \over (2mx^2)^2} \left( 
dx^2 dx^2 + dx^\mu dx^\nu \hat g_{\mu\nu} \right). 
\label{metric}
\ee
The SO(2,2) invariant AdS metric corresponds to the case 
$\hat g_{\mu\nu} = \eta_{\mu\nu}$ but we consider the general 
$\hat g_{\mu\nu}$. We define $\hat e_\mu{}^a$ by 
$\hat g_{\mu\nu} = \hat e_\mu{}^a \hat e_\nu{}^b \eta_{ab}$ and 
\be
\psi_{\mu\pm}^{\alpha\dot\alpha} = (2mx^2)^{\mp{1 \over 2}} 
\hat\psi_{\mu\pm}^{\alpha\dot\alpha}, \quad
\psi_{\mu\pm}^{\alpha'\dot\alpha'} = (2mx^2)^{\pm{1 \over 2}} 
\hat\psi_{\mu\pm}^{\alpha'\dot\alpha'}, \quad
A_\mu^\alpha{}_\beta = \hat A_\mu^\alpha{}_\beta, \quad
A_\mu^{\alpha'}{}_{\beta'} = \hat A_\mu^{\alpha'}{}_{\beta'}, 
\label{hatfield}
\ee
where $\psi_{\mu\pm} = {1 \over 2} ( 1 \pm \gamma^2 ) \psi_\mu$. 
As discussed in ref.\ \cite{NT} the field equations show that 
the fields with $\hat{\ }$ behave as ${\cal O}((x^2)^0)$ near 
the boundary. 
We impose boundary conditions on the fields at $x^2=0$ as 
\ba
\A\A \hat e_\mu{}^a = e_{0\mu}{}^a, \quad
\hat\psi_{\mu +}^{\alpha\dot\alpha} 
= \psi_{0\mu +}^{\alpha\dot\alpha}, \quad
\hat\psi_{\mu -}^{\alpha'\dot\alpha'} 
= \psi_{0\mu -}^{\alpha'\dot\alpha'}, \nonu
\A\A \hat A_\mu^{(+)\alpha}{}_\beta 
= A_{0\mu}^{(+)\alpha}{}_\beta, \quad
\hat A_\mu^{(-)\alpha'}{}_{\beta'} 
= A_{0\mu}^{(-)\alpha'}{}_{\beta'}, 
\label{bbehavior}
\ea
where $\hat A_\mu^{(\pm)} = {1 \over 2} 
( \hat g_{\mu\nu} \pm \hat e \epsilon_{\mu\nu} ) \hat A^\nu$. 
The fields $\phi_{0I} = ( e_{0\mu}{}^a, 
\psi_{0\mu+}^{\alpha\dot\alpha}, \psi_{0\mu-}^{\alpha'\dot\alpha'}, 
A_{0\mu}^{(+)\alpha}{}_\beta, A_{0\mu}^{(-)\alpha'}{}_{\beta'} )$ 
are fixed functions on the boundary. 
Since field equations of the Rarita-Schwinger fields and 
the gauge fields are first order, boundary conditions are 
imposed on only half of their components. 
Other components of the fields on the boundary 
are non-local functionals of the fields $\phi_{0I}$, 
which are obtained by solving the field equations. 
We also introduce notations $\psi_{0\mu-}^{\alpha\dot\alpha}$, 
$\psi_{0\mu+}^{\alpha'\dot\alpha'}$, 
$A_{0\mu}^{(-)\alpha}{}_\beta$, $A_{0\mu}^{(+)\alpha'}{}_{\beta'}$ 
defined by $\hat\psi_{\mu-}^{\alpha\dot\alpha} \rightarrow 
\psi_{0\mu-}^{\alpha\dot\alpha}$, etc. 
\par
The parameters of the residual symmetries, which preserve 
the gauge conditions (\ref{gaugefix}), near the 
boundary $x^2 = 0$ are 
\ba
\xi^2 \A = \A - x^2 \Lambda_0(x^0, x^1), \qquad
\xi^\mu = \xi_0^\mu(x^0, x^1) + {\cal O}((x^2)^2), \nonu
\lambda^{ab} \A = \A \lambda_0^{ab}(x^0, x^1) + {\cal O}(x^2), \qquad
\lambda^{a2} = {\cal O}(x^2), \nonu
\epsilon_\pm^{\alpha\dot\alpha} 
\A = \A (2mx^2)^{\mp{1 \over 2}} \left[ 
\epsilon_{0\pm}^{\alpha\dot\alpha}(x^0, x^1) 
+ {\cal O}(x^2) \right], \nonu
\epsilon_\pm^{\alpha'\dot\alpha'} 
\A = \A (2mx^2)^{\pm{1 \over 2}} \left[ 
\epsilon_{0\pm}^{\alpha'\dot\alpha'}(x^0, x^1) 
+ {\cal O}(x^2) \right], \nonu
\theta^\alpha{}_\beta 
\A = \A \theta_0^\alpha{}_\beta(x^0, x^1) 
+ {\cal O}((x^2)^2), \qquad
\theta^{\alpha'}{}_{\beta'} 
= \theta_0^{\alpha'}{}_{\beta'}(x^0, x^1) + {\cal O}((x^2)^2), 
\label{residual}
\ea
where $\Lambda_0$, $\xi_0^\mu$, $\lambda_0^{ab}$, 
$\epsilon_{0\pm}^{\alpha\dot\alpha}$, 
$\epsilon_{0\pm}^{\alpha'\dot\alpha'}$, $\theta_0^\alpha{}_\beta$ 
and $\theta_0^{\alpha'}{}_{\beta'}$ are arbitrary functions of 
$x^0$ and $x^1$. Order ${\cal O}(x^2)$ and ${\cal O}((x^2)^2)$ 
terms are non-local functionals of these functions and the 
fields $\phi_{0I}$. Substituting eqs.\ (\ref{hatfield}), 
(\ref{bbehavior}), (\ref{residual}) into eq.\ (\ref{localsym}) and 
taking the limit $x^2 \rightarrow 0$ we obtain transformations of 
the fields on the boundary $\phi_{0I}$. These transformations are 
shown to coincide with those of the two-dimensional (4,4) conformal 
supergravity \cite{PvN}. 
\par
The two-dimensional (4,4) conformal supergravity \cite{PvN} 
contains a zweibein $\tilde e_\mu{}^a$, doubly-symplectic 
Majorana Rarita-Schwinger fields 
$\tilde\psi_\mu^{\alpha\dot\alpha}$ and SU(2) gauge fields 
$\tilde A_\mu^\alpha{}_\beta$. These fields are identified 
with the boundary fields $\phi_{0I}$ as 
\ba
\tilde e_\mu{}^a \A = \A e_{0\mu}{}^a, \qquad
\tilde \psi_\mu^{\alpha\dot\alpha} 
= \psi_{0\mu+}^{\alpha\dot\alpha} 
+ \psi_{0\mu-}^{\alpha'\dot\alpha'}, \qquad
\tilde A_\mu^\alpha{}_\beta
= A_{0\mu}^{(+)\alpha}{}_\beta 
+ A_{0\mu}^{(-)\alpha'}{}_{\beta'}, 
\label{44sugra}
\ea
where $\alpha' = \alpha$, $\dot\alpha' = \dot\alpha$, 
$\beta' = \beta$. 
We find that transformation laws of these fields derived from 
the three-dimensional transformations (\ref{localsym}) are 
\ba
\delta \tilde e_\mu{}^a \A = \A 
\tilde\xi^\nu \partial_\nu \tilde e_\mu{}^a 
+ \partial_\mu \tilde\xi^\nu \tilde e_\nu{}^a 
+ \tilde\Lambda e_\mu{}^a 
- \tilde\lambda^a{}_b \tilde e_\mu{}^b 
+ {1 \over 2} \, i \, \bar{\tilde\epsilon}_{\alpha\dot\alpha} 
\gamma^a \tilde\psi_\mu^{\alpha\dot\alpha}, \nonu
\delta \tilde\psi_\mu^{\alpha\dot\alpha} \A = \A 
\tilde\xi^\nu \partial_\nu \tilde\psi_\mu^{\alpha\dot\alpha} 
+ \partial_\mu \tilde\xi^\nu \tilde\psi_\nu^{\alpha\dot\alpha} 
+ {1 \over 2} \tilde\Lambda \tilde\psi_\mu^{\alpha\dot\alpha} 
- {1 \over 4} \tilde\lambda^{ab} \gamma_{ab} 
\tilde\psi_\mu^{\alpha\dot\alpha} \nonu
\A \A - \tilde\theta_{\rm V}^\alpha{}_\beta 
\tilde\psi_\mu^{\beta\dot\alpha} 
- \tilde\theta_{\rm A}^\alpha{}_\beta \gamma_2 
\tilde\psi_\mu^{\beta\dot\alpha} 
+ \tilde{\cal D}_\mu \tilde\epsilon^{\alpha\dot\alpha} 
+ \tilde\gamma_\mu \tilde\eta^{\alpha\dot\alpha}, \nonu
\delta \tilde A_\mu^i 
\A = \A 
\tilde\xi^\nu \partial_\nu A_\mu^i 
+ \partial_\mu \tilde\xi^\nu \tilde A_\nu^i 
+ \tilde {\cal D}_\mu \tilde\theta_{\rm V}^i 
+ \tilde e \epsilon_{\mu\nu} \tilde {\cal D}^\nu 
\tilde\theta_{\rm A}^i \nonu
\A \A + \left( {1 \over 4} 
\bar{\tilde\epsilon}_{\alpha\dot\alpha} 
\tilde\gamma_\mu \tilde\gamma^{\rho\sigma} 
\tilde\psi_{\rho\sigma}^{\beta\dot\alpha} 
+ {1 \over 2} \bar{\tilde\eta}_{\alpha\dot\alpha} \tilde\gamma^\nu 
\tilde\gamma_\mu \tilde\psi_\nu^{\beta\dot\alpha} \right) 
( \sigma^i )^\alpha{}_\beta, 
\label{44trans}
\ea
where $\tilde{\cal D}_\mu$ is the covariant derivative including 
the gauge field $\tilde A_\mu^\alpha{}_\beta$ and 
$\tilde\psi_{\mu\nu}^{\alpha\dot\alpha}$ is the field strengths 
of the Rarita-Schwinger fields 
\ba
\tilde{\cal D}_\mu \tilde\epsilon^{\alpha\dot\alpha} 
\A = \A \tilde D_\mu \tilde\epsilon^{\alpha\dot\alpha} 
+ \tilde A_\mu^\alpha{}_\beta \tilde\epsilon^{\beta\dot\alpha}, \nonu
\tilde\psi_{\mu\nu}^{\alpha\dot\alpha} 
\A = \A \tilde{\cal D}_\mu \tilde\psi_\nu^{\alpha\dot\alpha} 
- \tilde{\cal D}_\nu \tilde\psi_\mu^{\alpha\dot\alpha}. 
\label{44def}
\ea
The transformation parameters with a tilde in eq.\ (\ref{44trans}) 
are related to those in eq.\ (\ref{residual}) as 
\ba
\tilde\xi^\mu \A = \A \xi^\mu, \quad
\tilde\Lambda = \Lambda_0, \quad
\tilde\lambda^{ab} = \lambda_0^{ab}, \quad
\tilde\epsilon^{\alpha\dot\alpha} 
= \epsilon_{0+}^{\alpha\dot\alpha} 
+ \epsilon_{0-}^{\alpha'\dot\alpha'}, \nonu
\tilde\eta^{\alpha\dot\alpha} 
\A = \A 2m ( \epsilon_{0-}^{\alpha\dot\alpha} 
- \epsilon_{0+}^{\alpha'\dot\alpha'} ) 
+ {1 \over 2} ( \slA_0^\alpha{}_\beta 
- \slA_0^{\alpha'}{}_{\beta'} ) \gamma_2 
\tilde\epsilon^{\beta\dot\alpha}, \nonu
\tilde\theta_{\rm V}^\alpha{}_\beta 
\A = \A {1 \over 2} \left( \theta_0^\alpha{}_\beta 
+ \theta_0^{\alpha'}{}_{\beta'} \right), \quad
\tilde\theta_{\rm A}^\alpha{}_\beta 
= {1 \over 2} \left( \theta_0^\alpha{}_\beta 
- \theta_0^{\alpha'}{}_{\beta'} \right), 
\label{44parameter}
\ea
where $\alpha' = \alpha$, $\dot\alpha' = \dot\alpha$, 
$\beta' = \beta$. 
The transformations (\ref{44trans}) represent 
general coordinate ($\tilde\xi^\mu$), local Lorentz 
($\tilde\lambda^{ab}$), Weyl ($\tilde\Lambda$), 
vector SU(2) gauge ($\theta_{\rm V}^\alpha{}_\beta$), 
axial vector gauge ($\theta_{\rm A}^\alpha{}_\beta$), 
local super ($\tilde\epsilon^{\alpha\dot\alpha}$) and 
super Weyl ($\tilde\eta^{\alpha\dot\alpha}$) transformations. 
These are local symmetry transformations of the (4,4) conformal 
supergravity \cite{PvN}. 
\par
%
%
\newsection{Boundary terms and counterterms}
To compute anomalies (\ref{anomaly}) for the local transformations 
(\ref{44trans}) we need the total action (\ref{totalaction}) 
including the boundary terms and the local counterterms. 
The boundary terms are chosen such that the variational 
principle is well-defined \cite{HS,AF,HENNEAUX}. 
When we take a variation of the action, we obtain boundary 
terms arising from partial integrations in addition to bulk 
terms proportional to field equations. Such boundary terms 
should be cancelled by a variation of $S_{\rm boundary}$ 
in eq.\ (\ref{totalaction}) when the boundary conditions 
(\ref{bbehavior}) are used. 
We find that appropriate boundary terms are 
\ba
S_{\rm boundary}
\A = \A \int d^2 x \biggl[ 
{1 \over 2} \epsilon^{\mu\nu} \epsilon_{ABC} 
e_\mu{}^A \omega_\nu^{BC}
-{1 \over 2} i \epsilon^{\mu\nu} \left( 
\bar\psi_{\mu+\alpha\dot\alpha} \psi_{\nu-}^{\alpha\dot\alpha} 
+ \bar\psi_{\mu-\alpha'\dot\alpha'} 
\psi_{\nu+}^{\alpha'\dot\alpha'} \right) \nonu
\A \A - {1 \over 8m} \hat e {\hat g}^{\mu\nu} 
\tr \left( A_\mu A_\nu + A'_\mu A'_\nu \right) \biggr]. 
\label{boundaryaction}
\ea
The first term is equivalent to the usual boundary term for the 
Einstein action, which is  proportional to the trace of the 
second fundamental form of the boundary \cite{GH}. 
\par
The bulk action and the boundary terms are both divergent. 
One has to regularize them and subtract singular terms by 
the counterterms $\Delta S$. The singular terms turn out to 
be local functionals of the fields $\phi_{0I}$. 
We regularize divergences by restricting the range of $x^2$ 
to $x^2 > \epsilon$, where $\epsilon > 0$ is 
a cut-off parameter. The boundary is now at $x^2 = \epsilon$. 
The regularization is removed for $\epsilon \rightarrow 0$. 
\par
Let us evaluate singular terms in $S_{\rm bulk}$ and 
$S_{\rm boundary}$ to obtain the local counterterms $\Delta S$. 
Using the field equations the bulk action becomes 
\be
S_{\rm bulk} = \int d^2 x \int_\epsilon^\infty d x^2 
\left[ - {1 \over m (x^2)^3} \, \hat e 
- {1 \over 4 x^2} \, i \hat e \left( 
\bar{\hat\psi}_{\mu\alpha\dot\alpha} \hat\gamma^{\mu\nu} 
\hat\psi_\nu^{\alpha\dot\alpha} 
- \bar{\hat\psi}_{\mu\alpha'\dot\alpha'} \hat\gamma^{\mu\nu} 
\hat\psi_\nu^{\alpha'\dot\alpha'} \right) \right]. 
\label{bulkaction2}
\ee
As in ref.\ \cite{HS2} we then expand the metric in $x^2$ as 
\be
\hat g_{\mu\nu} = g_{0\mu\nu} + (2mx^2)^2 g_{(2)\mu\nu} 
+ (2mx^2)^2 \log(2mx^2)^2 h_{(2)\mu\nu} + {\cal O}((x^2)^3). 
\label{gexp}
\ee
Coefficients of these expansion $g_{(2)\mu\nu}$, $h_{(2)\mu\nu}$, 
$\cdots$ are determined as functions of the fields $\phi_{0I}$ 
by solving the field equations. 
To compute singular terms in eq.\ (\ref{bulkaction2}) 
we need to know the expansion of $\hat e = \sqrt{-\hat g}$, 
which are determined by $g_0^{\mu\nu} g_{(2)\mu\nu}$ and 
$g_0^{\mu\nu} h_{(2)\mu\nu}$. They can be obtained from 
the (22) component of the Einstein equation as 
in ref.\ \cite{HS2}. We find $g_0^{\mu\nu} h_{(2)\mu\nu} = 0$ and 
\be
g_0^{\mu\nu} g_{(2)\mu\nu} 
= - {1 \over 8m^2} \left[ 
R_0 + i m \left( 
\bar\psi_{0\mu\alpha\dot\alpha} \gamma_0^{\mu\nu} 
\psi_{0\nu}^{\alpha\dot\alpha} 
- \bar\psi_{0\mu\alpha'\dot\alpha'} \gamma_0^{\mu\nu} 
\psi_{0\nu}^{\alpha'\dot\alpha'} 
\right) \right]. 
\label{g2}
\ee
Then we find singular terms in eq.\ (\ref{bulkaction2}) as 
\be
S_{\rm bulk}^{\rm (div)} 
= \int d^2 x \, e_0 \left[ - {1 \over 2m\epsilon^2} 
- {1 \over 4m} \log\epsilon \, R_0 \right]. 
\label{bulkdiv}
\ee
The singularities of $\psi$ terms in eq.\ (\ref{bulkaction2}) 
have been cancelled by those coming from eq.\ (\ref{g2}). 
To obtain singular terms in the boundary terms 
(\ref{boundaryaction}) we first note that all the terms 
in eq.\ (\ref{boundaryaction}) 
except the first are finite. By using the explicit form of 
the spin connection the first term can be written as 
\ba
\int d^2 x \, {1 \over 2} \epsilon^{\mu\nu} \epsilon_{ABC} 
e_\mu{}^A \omega_\nu^{BC}
\A = \A \int d^2 x \left[ {1 \over m \epsilon^2} \hat e 
- {1 \over 2m \epsilon} \partial_2 \hat e \right] \nonu
\A = \A \int d^2 x \, {1 \over m \epsilon^2} e_0 
+ {\cal O}(\epsilon), 
\label{boundarydiv}
\ea
where we have used eqs.\ (\ref{gexp}), (\ref{g2}) in the second 
equality. The counterterms are chosen to cancel the singularities 
in eqs.\ (\ref{bulkdiv}), (\ref{boundarydiv}) as 
\be
\Delta S = \int d^2 x \left[ 
- {1 \over 2 m \epsilon^2} e_0 
+ {1 \over 4 m} \log\epsilon \, e_0 R_0 \right] 
+ \Delta S_{\rm finite}, 
\label{counterterm}
\ee
where $\Delta S_{\rm finite}$ is a finite local term. 
The second term of the integrand is proportional to the Euler 
density, which is a total derivative locally, and will be 
ignored in the following. 
\par
We note that $\Delta S$ is a local functional of the boundary 
fields $\phi_{0I}$. In principle, it could depend on other 
components of the fields such as $\psi_{0\mu-}^{\alpha\dot\alpha}$, 
which is a non-local functional of $\phi_{0I}$. From the AdS point of 
view it is not clear why they should be local. The locality of 
singular terms may be understood by the IR-UV connection \cite{SW}. 
Infrared divergences in the bulk theory can be interpreted as 
ultraviolet divergences in the boundary CFT. 
Since ultraviolet divergences are short distance property of 
the CFT, they should be local. 
\par
%
%
\newsection{Anomalies}
We compute a variation of the total action (\ref{totalaction}) 
under the local symmetry transformations (\ref{44trans}) 
and obtain anomalies of the two-dimensional (4,4) CFT. 
Since the boundary transformations (\ref{44trans}) 
are generated from the bulk transformations (\ref{localsym}), 
we can use the latter to compute a variation in the bulk. 
Let us first consider the bosonic transformations. For the moment 
we assume $\Delta S_{\rm finite} = 0$ in eq.\ (\ref{counterterm}). 
Under the three-dimensional general coordinate transformation 
with a parameter $\xi^M$ the bulk action transforms as 
\ba
\delta_G S_{\rm bulk} 
\A = \A \int d^2 x \int_\epsilon^\infty d x^2 \, \partial_M 
\left( \xi^M {\cal L}_{\rm bulk} \right) \nonu
\A = \A \int d^2 x \, \epsilon \Lambda_0 
{\cal L}_{\rm bulk} \nonu
\A = \A \int d^2 x \, e_0 \Lambda_0 \left[ 
- {1 \over m\epsilon^2} + {1 \over 4m} R_0 \right], 
\ea
where ${\cal L}_{\rm bulk}$ is the integrand of 
eq.\ (\ref{bulkaction}) or eq.\ (\ref{bulkaction2}). 
In the last equality we have used eq.\ (\ref{g2}). 
The transformation of the boundary terms (\ref{boundaryaction}) 
and the local counterterms (\ref{counterterm}) can be easily 
computed by using the transformation laws (\ref{44trans}) 
and eq.\ (\ref{boundarydiv}). Using the quantities in 
eqs.\ (\ref{44sugra}), (\ref{44parameter}) we obtain the 
transformation of the total action as 
\be
\delta_G S = {1 \over 4m} \int d^2 x \, 
\tilde e \tilde \Lambda \tilde R. 
\label{wanomaly}
\ee
We see that the total action is invariant under the 
two-dimensional general coordinate transformation 
but not under the Weyl transformation. 
As for the local Lorentz transformation the bulk action 
(\ref{bulkaction}) and the local counterterms (\ref{counterterm}) 
are manifestly invariant. The boundary action 
(\ref{boundaryaction}) is also invariant when the first term 
is expressed as in eq.\ (\ref{boundarydiv}). 
Therefore, the total action is invariant under the local 
Lorentz transformation
\be
\delta_L S = 0. 
\label{lanomaly}
\ee
Under the SU(2) $\times$ SU(2) gauge transformation 
the Chern-Simons terms in the bulk action (\ref{bulkaction}) 
give two-dimensional chiral gauge anomaly \cite{ZWZ}. 
Combining with the transformation of the boundary action we obtain 
\be
\delta_g S 
= {1 \over 2m} \int d^2 x \, \epsilon^{\mu\nu} 
\tr \left( \theta_0 \partial_\mu A_{0\nu}^{(+)} 
- \theta'_0 \partial_\mu A_{0\nu}^{'(-)} \right). 
\label{ganomaly}
\ee
It is possible to preserve the diagonal vector SU(2) subgroup 
of the SU(2) $\times$ SU(2) gauge symmetry by introducing an 
appropriate finite counterterm. By choosing 
\be
\Delta S_{\rm finite} = - {1 \over 2m} \int d^2 x \, 
\epsilon^{\mu\nu} \tr \left( 
A_{0\mu}^{(+)} A_{0\nu}^{'(-)} \right) 
\label{finitect}
\ee
we obtain 
\be
\delta_g S 
= {1 \over 2m} \int d^2 x \, \epsilon^{\mu\nu} 
\tr \left( \tilde\theta_{\rm A} \tilde F_{\mu\nu} \right), 
\label{ganomaly2}
\ee
where $\tilde F_{\mu\nu}$ is the field strength of the SU(2) 
gauge field $\tilde A_\mu$. 
We see that there is no anomaly in the vector SU(2) gauge 
transformation with the parameter $\tilde\theta_{\rm V}$. 
Since the finite counterterm (\ref{finitect}) is invariant 
under the general coordinate and local Lorentz transformations, 
the introduction of it does not change the results 
(\ref{wanomaly}), (\ref{lanomaly}). 
\par
To obtain the supertransformation of the action we first compute 
the variation of the total action as 
\ba
\delta_Q S 
\A = \A {1 \over 2m} \int d^2 x \biggl[ 
{1 \over \epsilon} \hat e \hat e_a{}^\mu 
\delta_Q \omega_\mu{}^{a2} 
+ {1 \over \epsilon^2} \delta_Q e_0 \nonu
\A \A - 2im \epsilon^{\mu\nu} \left( 
\delta_Q \bar\psi_{0\mu+\alpha\dot\alpha} 
\psi_{0\nu-}^{\alpha\dot\alpha} 
+ \delta_Q \bar\psi_{0\mu-\alpha'\dot\alpha'} 
\psi_{0\nu+}^{\alpha'\dot\alpha'} \right) \nonu
\A \A - e_0 \tr \left( A_0^{(-)\mu} \delta_Q A_{0\mu} 
+ A_0^{'(+)\mu} \delta_Q A'_{0\mu} \right) \nonu
\A \A - {1 \over 4} \delta_Q \left( e_0 g_0^{\mu\nu} \right) 
\tr \left( A_{0\mu} A_{0\nu} + A'_{0\mu} A'_{0\nu} \right) 
\biggr] 
+ \delta_Q \Delta S_{\rm finite}. 
\label{qvariation}
\ea
The first two terms can be written as 
\ba
\A\A {1 \over \epsilon} \hat e \hat e_a{}^\mu 
\delta_Q \omega_\mu{}^{a2} 
+ {1 \over \epsilon^2} \delta_Q e_0 \nonu
\A\A \qquad = - {1 \over \epsilon} \hat e 
\delta_Q \hat e_a{}^\mu \bar\omega_\mu{}^{a2} 
+ {1 \over \epsilon} \hat e \delta_Q \left( 
\hat e^{-1} \partial_2 \hat e \right) 
- {1 \over \epsilon^2} \delta_Q \hat e 
+ {1 \over \epsilon^2} \delta_Q e_0 \nonu
\A\A \qquad = - {1 \over \epsilon} \hat e 
\delta_Q \hat e_a{}^\mu \bar\omega_\mu{}^{a2} 
+ 2m^2 \delta_Q \left( 
e_0 g_0^{\mu\nu} g_{(2)\mu\nu} \right) 
- 4m^2 \delta_Q e_0 g_0^{\mu\nu} g_{(2)\mu\nu}, 
\label{firsttwo}
\ea
where $\bar\omega_\mu{}^{a2} = \omega_\mu{}^{a2} + 2m e_\mu{}^a$ 
and we have used eq.\ (\ref{gexp}). 
The second term in the last equation does not contribute 
because of the formula 
\be
\delta_Q \int d^2 x \, e_0 g_0^{\mu\nu} g_{(2)\mu\nu} = 0. 
\ee
This can be shown by using the Rarita-Schwinger field equations 
expressed as 
\ba
\epsilon^{\mu\nu} {\cal D}_{0\mu} \psi_{0\nu-}^{\alpha\dot\alpha} 
\A = \A - {1 \over 4m\epsilon} \epsilon^{\mu\nu} 
\gamma_a \psi_{0\nu+}^{\alpha\dot\alpha} 
\bar\omega_\mu{}^{a2}, \nonu
\epsilon^{\mu\nu} {\cal D}_{0\mu} \psi_{0\nu+}^{\alpha'\dot\alpha'} 
\A = \A  {1 \over 4m\epsilon} \epsilon^{\mu\nu} \gamma_a 
\psi_{0\nu-}^{\alpha'\dot\alpha'} \bar\omega_\mu{}^{a2}, \nonu
\epsilon^{\mu\nu} {\cal D}_{0\mu} \psi_{0\nu+}^{\alpha\dot\alpha} 
\A = \A - 2m \epsilon^{\mu\nu} 
\gamma_{0\mu} \psi_{0\nu-}^{\alpha\dot\alpha}, \nonu
\epsilon^{\mu\nu} {\cal D}_{0\mu} \psi_{0\nu-}^{\alpha'\dot\alpha'} 
\A = \A 2m \epsilon^{\mu\nu} 
\gamma_{0\mu} \psi_{0\nu+}^{\alpha'\dot\alpha'}. 
\label{rseq}
\ea
Eq.\ (\ref{rseq}) also shows that the third term in 
eq.\ (\ref{qvariation}) can be written as 
\ba
\A\A {1 \over 2\epsilon} i \epsilon^{\mu\nu} \left( 
\bar\epsilon_{0+\alpha\dot\alpha} \gamma_{0\nu} 
\psi_{0a+}^{\alpha\dot\alpha} 
- \bar\epsilon_{0-\alpha'\dot\alpha'} \gamma_{0\nu} 
\psi_{0a-}^{\alpha'\dot\alpha'} 
- \bar\epsilon_{0+\alpha\dot\alpha} \gamma_a 
\psi_{0\nu+}^{\alpha\dot\alpha} 
+ \bar\epsilon_{0-\alpha'\dot\alpha'} \gamma_a 
\psi_{0\nu-}^{\alpha'\dot\alpha'} \right) 
\bar\omega_\mu{}^{a2} \nonu
\A\A \qquad\qquad + 4 i m^2 \epsilon^{\mu\nu} \left( 
\bar\epsilon_{0-\alpha\dot\alpha} \gamma_{0\mu} 
\psi_{0\nu-}^{\alpha\dot\alpha} 
- \bar\epsilon_{0+\alpha'\dot\alpha'} \gamma_{0\mu} 
\psi_{0\nu+}^{\alpha'\dot\alpha'} \right). 
\label{thirdterm}
\ea
The terms proportional to $\bar\omega_\mu{}^{a2}$ in this 
equation, in turn, can be written as 
\be
{1 \over 2\epsilon} i \epsilon^{\mu\nu} \left( 
\bar\epsilon_{0+\alpha\dot\alpha} \gamma_{0\nu} 
\psi_{0\mu+}^{\alpha\dot\alpha} 
- \bar\epsilon_{0-\alpha'\dot\alpha'} \gamma_{0\nu} 
\psi_{0\mu-}^{\alpha'\dot\alpha'} \right) 
\bar\omega_a{}^{a2}
= 4 m^2 \delta_Q e_0 g_0^{\mu\nu} g_{(2)\mu\nu}, 
\ee
which cancels the last term in eq.\ (\ref{firsttwo}). 
On the other hand, by using eq.\ (\ref{rseq}) the remaining 
terms in eq.\ (\ref{thirdterm}) become 
\be
- 2 i m \epsilon^{\mu\nu} \left[ 
\bar\epsilon_{0-\alpha\dot\alpha} 
{\cal D}_{0\mu} \psi_{0\nu+}^{\alpha\dot\alpha} 
+ \bar\epsilon_{0+\alpha'\dot\alpha'} 
{\cal D}_{0\mu} \psi_{0\nu-}^{\alpha'\dot\alpha'} 
\right]. 
\ee
Collecting these results and after some more algebra we finally 
obtain the supertransformation of the total action as 
\be
\delta_Q S = {1 \over 4m} i \int d^2 x \, \tilde e \, 
\bar{\tilde\eta}_{\alpha\dot\alpha} \tilde\gamma^{\mu\nu} 
\tilde\psi_{\mu\nu}^{\alpha\dot\alpha}, 
\label{qanomaly}
\ee
where $\tilde\psi_\mu^{\alpha\dot\alpha}$ is the field strength 
of the Rarita-Schwinger fields defined in eq.\ (\ref{44def}). 
\par
To summarize, our results of anomalies for the local symmetry 
transformations are given by eqs.\ (\ref{wanomaly}), 
(\ref{lanomaly}), (\ref{ganomaly2}) and (\ref{qanomaly}). 
There is no anomaly in two-dimensional general coordinate, 
local Lorentz, vector SU(2) gauge and local super 
transformations. The remaining symmetries, i.e., 
two-dimensional Weyl, axial-vector gauge and super Weyl 
symmetries, are anomalous. The anomalies (\ref{wanomaly}), 
(\ref{lanomaly}), (\ref{ganomaly2}) and (\ref{qanomaly}) form 
an anomaly supermultiplet \cite{GRISARU}, in the sence that 
they transform to one another by the two-dimensional local 
supertransformation. This can be shown by using the Wess-Zumino 
consistency condition \cite{WZ,TANII} and the absence of the 
two-dimensional local supersymmetry anomaly. 
These anomalies correspond to the (4,4) 
super Virasoro algebras with the central charge 
$c = {6\pi \over m} = {3l \over 2G}$. 
Anomalies in the ($p,q$) theories for other values of 
$p,q =1,2,4$ can be obtained from these results by truncation 
of the fields discussed at the end of sect.\ 2. 

\vspace{7mm}

We would like to thank M. Natsuume for informing us of 
ref.\ \cite{BK}. This work is supported in part by the 
Grant-in-Aid for Scientific Research on Priority Area 707 
``Supersymmetry and Unified Theory of Elementary Particles'', 
Japan Ministry of Education. 

%
%
\newcommand{\NP}[1]{{\it Nucl.\ Phys.\ }{\bf #1}}
\newcommand{\PL}[1]{{\it Phys.\ Lett.\ }{\bf #1}}
\newcommand{\CMP}[1]{{\it Commun.\ Math.\ Phys.\ }{\bf #1}}
\newcommand{\MPL}[1]{{\it Mod.\ Phys.\ Lett.\ }{\bf #1}}
\newcommand{\IJMP}[1]{{\it Int.\ J. Mod.\ Phys.\ }{\bf #1}}
\newcommand{\PR}[1]{{\it Phys.\ Rev.\ }{\bf #1}}
\newcommand{\PRL}[1]{{\it Phys.\ Rev.\ Lett.\ }{\bf #1}}
\newcommand{\PTP}[1]{{\it Prog.\ Theor.\ Phys.\ }{\bf #1}}
\newcommand{\PTPS}[1]{{\it Prog.\ Theor.\ Phys.\ Suppl.\ }{\bf #1}}
\newcommand{\AP}[1]{{\it Ann.\ Phys.\ }{\bf #1}}
\newcommand{\ATMP}[1]{{\it Adv.\ Theor.\ Math.\ Phys.\ }{\bf #1}}

\begin{thebibliography}{100}
%
\bibitem{MAL} J. Maldacena, 
        The large $N$ limit of superconformal field theories 
        and supergravity, \ATMP{2} (1998) 231, hep-th/9711200. 
\bibitem{GKP} S.S. Gubser, I.R. Klebanov and A.M. Polyakov, 
        Gauge theory correlators from noncritical string theory, 
        \PL{B428} (1998) 105, hep-th/9802109. 
\bibitem{WITTEN} E. Witten, 
        Anti de Sitter space and holography, 
        \ATMP{2} (1998) 253, hep-th/9802150. 
%
\bibitem{FFZ} S. Ferrara, C. Fr\o nsdal and A. Zaffaroni, 
        On $N=8$ supergravity on ${\rm AdS}_5$ and $N=4$ 
        superconformal Yang-Mills theory, \NP{B532} (1998) 153, 
        hep-th/9802203. 
\bibitem{LT} H. Liu and A.A. Tseytlin, 
        $D=4$ super Yang-Mills, $D=5$ gauged supergravity 
        and $D=4$ conformal supergravity, \NP{B533} (1998) 88, 
        hep-th/9804083. 
%
\bibitem{NT} M. Nishimura and Y. Tanii, Supersymmetry in the 
        AdS/CFT correspondence, \PL{B446} (1999) 37, hep-th/9810148. 
%
\bibitem{AT} A. Ach\'ucarro and P.K. Townsend, 
        A Chern-Simons action for three-dimensional anti-de 
        Sitter supergravity theories, \PL{B180} (1986) 89. 
%
\bibitem{HS2} M. Henningson and K. Skenderis, 
        The holographic Weyl anomaly, 
        {\it JHEP} {\bf 07} (1998) 023, hep-th/9806087. 
\bibitem{HKL} S. Hyun, W.T. Kim and J. Lee, 
        Statistical entropy and AdS/CFT correspondence in BTZ 
        black holes, \PR{D59} (1999) 084020, hep-th/9811005. 
\bibitem{NO} S. Nojiri and S.D. Odintsov, Conformal anomaly for 
        dilaton coupled theories from AdS/CFT correspondence, 
        \PL{B444} (1998) 92, hep-th/9810008; 
        On the conformal anomaly from higher derivative gravity in 
        AdS/CFT correspondence, hep-th/9903033. 
\bibitem{GW} C.R. Graham and E. Witten, Conformal anomaly of 
        submanifold observables in AdS/CFT correspondence, 
        hep-th/9901021. 
\bibitem{APTY} O. Aharony, J. Pawe\l czyk, S. Theisen and 
        S. Yankielowicz, A note on anomalies in the AdS/CFT 
        correspondence, hep-th/9901134. 
\bibitem{BK} V. Balasubramanian and P. Kraus, 
        A stress tensor for anti-de Sitter gravity, hep-th/9902121. 
%
\bibitem{PvN} M. Pernici and P. van Nieuwenhuizen, 
        A covariant action for the SU(2) spinning string as a 
        hyperk\"ahler or quaternionic nonlinear sigma model, 
        \PL{B169} (1986) 381. 
%
\bibitem{SV} A. Strominger and C. Vafa, Microscopic origin of the 
        Bekenstein-Hawking entropy, \PL{B379} (1996) 99, 
        hep-th/9601029. 
\bibitem{CM} C.G. Callan and J.M. Maldacena, D-brane approach to 
        black hole quantum mechanics, \NP{B472} (1996) 591, 
        hep-th/9602043. 
%
\bibitem{MS} J. Maldacena and A. Strominger, 
        ${\rm AdS}_3$ black holes and a stringy exclusion principle, 
        {\it JHEP} {\bf 12} (1998) 005, hep-th/9804085. 
\bibitem{BOER} J. de Boer, Six-dimensional supergravity on 
        ${\rm S}^3 \times {\rm AdS}_3$ and 2d conformal field theory, 
        hep-th/9806104. 
%
\bibitem{BH} J.D. Brown and M. Henneaux, 
        Central charges in the canonical realization of asymptotic 
        symmetries: an example from three dimensional gravity, 
        \CMP{104} (1986) 207. 
%
\bibitem{HS} M. Henningson and K. Sfetsos, Spinors and the AdS/CFT 
        correspondence, \PL{B431} (1998) 63, hep-th/9803251. 
\bibitem{AF} G.E. Arutyunov and S.A. Frolov, On the origin of 
        supergravity boundary terms in the AdS/CFT correspondence, 
        hep-th/9806216. 
\bibitem{HENNEAUX} M. Henneaux, Boundary terms in the AdS/CFT 
         correspondence for spinor fields, hep-th/9902137. 
\bibitem{GH} G.W. Gibbons and S.W. Hawking, Action integrals and 
        partition functions in quantum gravity, \PR{D15} (1977) 2752. 
%
\bibitem{SW} L. Susskind and E. Witten, The holographic bound 
        in anti-de Sitter space, hep-th/9805114. 
%
\bibitem{ZWZ} B. Zumino, Y.-S. Wu and A. Zee, Chiral anomalies, 
        higher dimensions, and differential geometry, 
        \NP{B239} (1984) 477. 
%
\bibitem{GRISARU} M.T. Grisaru, Anomalies in supersymmetric 
        theories, in {\it Recent Developments in Gravitation}, 
        eds.\ M. Levy and S. Deser (Plenum Press, 1979). 
\bibitem{WZ} J. Wess and B. Zumino, Consequences of anomalous Ward 
        identities, \PL{B37} (1971) 95. 
\bibitem{TANII} Y. Tanii, The structures of anomalies in 
        supersymmetric string theories, \NP{B289} (1987) 187. 
%
\end{thebibliography}
\end{document}